\documentclass{report}
\usepackage{graphicx}
\usepackage{bm} 
\usepackage{fancybox}
\usepackage{comment}
\usepackage{color}
\usepackage{ohm-ngc}

\begin{document}
\title{Reciprocal Altruism-based Cooperation in a Social Network Game}
\author{Masanori Takano, Kazuya Wada \& Ichiro Fukuda}{CyberAgent Inc.}
\E-mail{\{takano\_masanori,wada\_kazuya,fukuda\_ichiro\}@cyberagent.co.jp}

\begin{abstract}
Cooperative behaviors are common in humans and are fundamental to our society. Theoretical and experimental studies have modeled environments in which the behaviors of humans, or agents, have been restricted to analyze their social behavior. However, it is important that such studies are generalized to less restrictive environments to understand human society. Social network games (SNGs) provide a particularly powerful tool for the quantitative study of human behavior. In SNGs, numerous players can behave more freely than in the environments used in previous studies; moreover, their relationships include apparent conflicts of interest and every action can be recorded. We focused on reciprocal altruism, one of the mechanisms that generate cooperative behavior. This study aims to investigate cooperative behavior based on reciprocal altruism in a less restrictive environment. For this purpose, we analyzed the social behavior underlying such cooperative behavior in an SNG. We focused on a game scenario in which the relationship between the players was similar to that in the Leader game. We defined cooperative behaviors by constructing a payoff matrix in the scenario. The results showed that players maintained cooperative behavior based on reciprocal altruism, and cooperators received more advantages than noncooperators. We found that players constructed reciprocal relationships based on two types of interactions, cooperative behavior and unproductive communication.

\end{abstract}

\clearpage
\section{Introduction}

Cooperative behaviors are common in humans, and they are fundamental to our society\cite{Fehr2003,Smitha}. However, noncooperators obtain more advantages than cooperators during interactions, because cooperators tend to be exploited by noncooperators\cite{Axelrod}; thus, natural selection should favor noncooperators. Nevertheless, humans cooperate with each other; therefore, they must have acquired mechanisms that ensure cooperation during the evolutionary process\cite{Barkow1995}.

Since cooperators tend to be exploited by noncooperators, these evolutionary dynamics require structured interactions in which cooperators interact more frequently with cooperators and noncooperators interact more frequently with noncooperators. Thus, humans must have acquired mechanisms for assortment between cooperators and noncooperators during the evolutionary process. Five theoretical mechanisms have been proposed\cite{Nowak2006}: kin selection, direct reciprocity, indirect reciprocity, spatial selection, and multilevel selection. Theoretical and experimental studies have presented evidence of these mechanisms\cite{Rand2013}. Evidence has been acquired using modeled environments with constrained behaviors of humans, or agents, to analyze their social behavior explicitly. However, it is important that this evidence be generalized to a less restrictive environment to understand human society\cite{Rand2013}.

Interactive online games are particularly powerful tools for the quantitative study of human society\cite{Castronova2006,Bainbridge2007,Szell2010,Szell2012,Szell2013}. In online games, numerous players can behave more freely than is possible in the environments used in the theoretical and experimental studies \cite{Nowak2006,Rand2013}, i.e., they do not need to select from a sequence of several alternatives, because they always have multiple alternatives. In addition, the actions of all players can be recorded. In the present study, we analyzed a social network game (SNG), such as Rage of Bahamut,\footnote{http://mobage.com/games/rage-of-bahamut} or Girl Friend BETA,\footnote{http://vcard.ameba.jp} to understand cooperative behavior among humans, because the following features of SNGs provide easier analysis of cooperative behavior. SNGs allow real players to cooperate and compete with others in situations when the player's benefit is represented by a quantitative value, such as a payoff in game theory.

We focused on cooperation based on direct reciprocity (reciprocal altruism) \cite{trivers1971}. The mechanism is a behavior whereby an individual acts in a manner that temporarily reduces its fitness, while increasing another individual's fitness, with the expectation that the other individual will behave in a similar manner at a later time. This behavior has been observed in humans\cite{Grujic2010,Grujic2012,Rand2011} and other primates\cite{PACKER1977}. In addition, the possibility of this behavior has been suggested even in vampire bats\cite{Wilkinson1990} and fishes\cite{Bshary2006}. Evolutionary game theory studies have also shown that reciprocal altruism drives the evolution of cooperation\cite{Lindgren1991,Nowak1993,Axelrod}.

Cooperation based on reciprocal altruism requires the following three conditions: 1) a long-term social relationship between individuals, 2) the capacity for individual recognition and memory of others' behavior, and 3) greater benefits from cooperation than costs of cooperation. If these conditions are satisfied, then an individual can judge whether interactions with another individual will provide benefits. Thus, cooperators increase their future benefits by cooperating with reciprocal cooperators. Simultaneously, it is difficult for noncooperators to interact with cooperators, because cooperators tend to select cooperators as interaction partners\cite{hauert2002}. Consequently, this may work as punishment inhibiting defective behavior by reciprocal cooperators with noncooperators\cite{Rand2009}.

This study aims to investigate reciprocal altruism that generates cooperative behavior in a less restrictive environment. For this purpose, we analyzed the effects of reciprocal altruism in an SNG, in which players could behave more freely than was possible in the environments used in the theoretical and experimental studies. Players competed and cooperated with each other in the SNG. In such an environment, players (i.e., humans) obviously have the capacity for individual recognition and the memory of others. Therefore, cooperation based on reciprocal altruism is expected to emerge in an SNG, because the above conditions are met. In this paper, we first confirmed the presence of cooperation based on reciprocal altruism. Then, we analyzed its effects to determine their benefits, and explored factors that reinforced cooperative behavior.

\section{Materials and Methods}

In this section, we provide the minimal SNG information and the definition of cooperative behavior in an SNG (see appendices \S1, \S2, and \S3 for game information, rules, and definition, respectively).

We analyzed cooperative behavior in the SNG, ``Girl Friend BETA,'' in which players acquired ``event points'' and competed in the rankings based on those points, because the players received better awards as their rankings increased. The player's ranking order was determined by the sum of event points obtained in the period from 3/25/2013 to 4/8/2013. 
It was impossible to analyze societal dynamics in this SNG, because the rules changed frequently. The situation in the SNG was also unstable in the early stage of this period; therefore, for simplification, we used only the data from the final three days.

The event points for players' actions correlate significantly with their levels, one of a player's attributes\footnote{Accurately, the event points per player's action depend primarily on players' attack power, which strongly correlates with their levels. The game did not store to a log file, hence we used players' levels as alternatives.}. Players must spend their energy to obtain event points; therefore, the number of players' actions is finite. There are two methods for replenishing these points, waiting for the points to replenish over time and using a paid item. Let ``payment amount'' be the sum of money spent by each player during the analysis period. Players must use their resources (items and time) effectively to progress to a higher ranking, because any player's time and money are finite.

Players belong to groups limited to 1–-50 players. The SNG is designed to ensure that cooperation with group members results in an effective game play. Players can communicate at any time through simple text messaging. This does not negatively affect either senders or receivers; nevertheless, its positive effects are also few\footnote{Players acquire a few points for a lottery that provides a card, when the players send messages to other at the beginning of each day. However, players must pay $200$ points for the lottery, and the effect of the card is small, i.e., the points do not increase players' abilities.}.
We targeted groups of five or more active players who logged in at least one or more times to analyze social interactions. In addtion, we limited data to intragroup communication and cooperation.

\begin{table}[hbtp]
 \caption{
 Payoff matrix of leader game, where $S+T>2R$ and $T>S>R>P$, 
 i.e., Pareto efficiency is achieved, when one cooperates and the other does not cooperate. 
 Then the cooperator obtains $S$, and the noncooperator $T$.}
 \label{table_leader_payoff_matrix}
 \begin{center}
  \begin{tabular}{c||c|c}
    & Cooperation & Noncooperation \\
   \hline \hline
   Cooperation & $R, R$ & $S, T$ \\
   \hline
   Noncooperation & $T, S$ & $P, P$ \\
  \end{tabular}
 \end{center}
\end{table}

We analyzed cooperative behavior based on reciprocal altruism in the above environment. It was difficult to track all cooperative behavior, because players can exhibit various behaviors in the SNG. Hence, we selected a specific cooperative behavior among various cooperative behaviors and regarded the frequency of that behavior as a measure of a player's cooperativeness.

We focused on a game scenario in which the relationship between players was similar to that in the Leader game (Table \ref{table_leader_payoff_matrix}), but it was not possible for both players to cooperate at the same time in this scenario (see appendix \S3). In the Leader game, Pareto efficiency is achieved, when one player cooperates and the other does not. Then the cooperator receives $S$, and the noncooperator $T$. That is, players receive a high payoff by sharing $S$ and $T$ on repeated plays of the game, a process known as $ST$ reciprocity\cite{Tanimoto2007}. We recognized this cooperative behavior, which provides the payoff $S$ to the other, as a cooperative behavior in this scenario.

\section{Results}

First, we confirm the presence of cooperation based on reciprocal altruism. We define the degree of reciprocity between player $i$ and player $j$ for this analysis as follows:
\begin{eqnarray}
r_{ij} = {\rm min}(C_{ij}, C_{ji}) / {\rm max}(C_{ij}, C_{ji}), \label{eq_recip_degree}
\end{eqnarray}
where $C_{ij}$ are the number of cooperative behaviors from player $i$ to player $j$, $C_{ji}$ are the number of cooperative behaviors from player $j$ to player $i$, i.e., $r_{ij}$ is asymptotic to one, when they mutually cooperate, and to zero, when one cooperates with the other and the other does not cooperate in return. $r_{ij}$ was observed, when $C_{ij} > 0$ or $C_{ji} > 0$.

\begin{figure}[h]
 \begin{center} 
  \includegraphics[width=70mm]{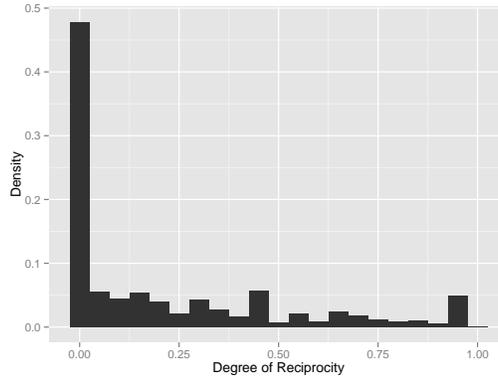}
  \caption{Density distribution of degree of reciprocity $r_{ij}$.}
  \label{fig_recip_degree_hist}
 \end{center}
\end{figure}

Figure \ref{fig_recip_degree_hist} shows the density distribution of $r_{ij}$. The mode was $0.0$ (i.e., nonreciprocal relationship). However, there were also some large values of $r_{ij}$ (i.e., reciprocal relationship). We regarded this relationships as reciprocal when $r_{ij} > 0.0$.

Second, we describe the effects of reciprocal relationships in regard to the benefits. The number of reciprocal relationships $f_i$ is expected to affect the number of event points $p_i$ positively, if cooperation based on reciprocity is effective in the SNG. Consider the following generalized linear model (GLM) to analyze the effect.
\begin{eqnarray}
p_i &\sim& {\rm NB}(r_i), \label{eq_pt_nb} \\
\ln r_i &=& \beta_1 \ln s_i + \beta_2 L_i + \beta_3 f_i + \beta_4. \nonumber 
\end{eqnarray}
This model is intended to explain the event points $p_i$ of player $i$ by player's level $L_i$ and the number of reciprocal relationships $f_i$. We also use the log of the payment amount for items $s_i$, because paid items proportionally increase the number of actions intended to obtain event points. ${\rm NB}(x)$ shows that $x$ follows a negative binomial distribution. We used a log link function and estimated its parameters using the maximum likelihood method with $5,000$ players whose $s_i > 0$, sampled at random. In the following analysis, we used the same link function and the same method of estimating parameters. 
We considered this model and a generalized linear mixed-effects model (GLMM) with a Poisson distribution, because the data showed over-dispersion when the GLM with a Poisson distribution was applied, then Akaike information criterion (AIC) selected this model.

\begin{table}
\caption{
The results of the regression analysis of event points $p_i$. $***$, $**$, $*$ indicate that the signs of regression coefficients did not change in a Wald-type $99.9\%$, $99\%$, and $95\%$ confidence interval (the symbols show the same meaning in the following figures).}
\label{tbl_pt_nb}
\begin{center}
  \begin{tabular}{l|rl}
    Explanatory Variable & Regression Coefficient &（Standard Error） \\ \hline \hline
    $\ln s_i$    & $ 0.3481788$ & $ (0.0060875) ^{***}$ \\ \hline
    $L_i$        & $ 0.0275284$ & $ (0.0003245) ^{***}$ \\ \hline
    $f_i$        & $ 0.1523005$ & $ (0.0080843) ^{***}$ \\ \hline
    Intercept    & $ 10.004722$ & $ (0.0411644) ^{***}$ \\ \hline
  \end{tabular}
\end{center}
\end{table}

Table \ref{tbl_pt_nb} shows the analysis results of the model. The regression coefficient of $f_i$ was positive, even after controlling for $s_i$ and $L_i$, i.e., a player with many reciprocal relationships received more event points than others with the same $s_i$ and $L_i$. Thus, reciprocal relationships increased the efficiency of payment, i.e., the relationships increased players' benefits.

Third, we report factors that drove the reciprocal relationships. Consider the following GLMM to analyze the factors:
\begin{eqnarray}
C_{ij} &\sim& {\rm Poisson}(\lambda_{ij}), \label{eq_l2r_glmmml}\\
\ln \lambda_{ij} &=& \beta_1 \ln C'_i C_j + \beta_2 C_{ji} + \beta_3 g_{ij} + \beta_4 g_{ji}  \nonumber \\
                  & & + \beta_5 L_i + \beta_6 L_j + \beta_7 m_i + \beta_8 + \sigma_i. \nonumber
\end{eqnarray}
This model is intended to explain the number of cooperative behaviors from $i$ to $j$ ($C_{ij}$) by the number of cooperative behaviors from $j$ to $i$ ($C_{ji}$), the frequency of messaging $g_{ij}$ and $g_{ji}$, both players' levels $L_i$ and $L_j$, their group size $m_i$, and the random effects of their group $\sigma_i$. In addition, we use the log of the product of $C'_i$ and $C_j$, which respectively shows the total number of cooperative behaviors of others with $i$ and the total number of cooperative behaviors $j$ with others, because this value is expected to increase $C_{ij}$ proportionally, if players cooperate at random. That is, this model estimates the effects of these explanatory variables on $C_{ij}$ comparing the random behavior. ${\rm Poisson}(x)$ shows that $x$ follows a Poisson distribution. We estimated its parameters with $30,000$ relationships between players whose $C'_i, C_j > 0$, sampled at random. 
We considered this model and a GLM with a negative binomial distribution, because the data showed over-dispersion when the GLM with a Poisson distribution was applied, then AIC selected this model.

\begin{table}
\caption{Results of regression analysis for cooperation between players.}
\label{tbl_l2r_glmmml}
\begin{center}
\begin{tabular}{l|rl}
  Explanatory Variable & Regression Coefficient &（Standard Error） \\ \hline \hline
  $\ln C'_i C_j$  & $ 0.6210306$ & $ (0.0041205) ^{***}$ \\ \hline
  $C_{ji}$        & $ 0.0122157$ & $ (0.0006525) ^{***}$ \\ \hline
  $g_{ij}$        & $-0.0008443$ & $ (0.0004852) $ \\ \hline
  $g_{ji}$        & $ 0.0021324$ & $ (0.0004206) ^{***}$ \\ \hline
  $L_i$           & $ 0.0004757$ & $ (0.0002047) ^{*}$ \\ \hline
  $L_j$           & $-0.0019128$ & $ (0.0001950) ^{***}$ \\ \hline 
  $m_i$           & $-0.1066716$ & $ (0.0022965) ^{***}$ \\ \hline
  Intercept       & $-0.1003971$ & $ (0.0283909) ^{***}$ \\ \hline
\end{tabular}
\end{center}
\end{table}

Table \ref{tbl_l2r_glmmml} shows the analysis results of the model. The regression coefficient of $C_{ji}$ and $g_{ji}$ were positive, even after controlling for $C_j$, $C'_i$, $L_i$ and $L_j$, i.e., to receive cooperation from others, it was important for players to cooperate with each other and to receive communications from others. In addition, this trend was stronger with smaller groups, because $m_i$ was negative.

Finally, we analyzed the effects of players' social relationships on their social behavior in win–-win situations, in which responding to a ``help'' request from
group members provides benefits to both the helper and the helped players (see 
appendix \S4). Then, the helper can obtain benefits even by helping someone. We
estimated the social relationships effect on whether a player who receives a help
request responds to the request. Consider the following GLM to analyze the effect of reciprocal relationships:
\begin{eqnarray}
H_{ij} &\sim& {\rm NB}(r_{ij}), \label{eq_l2r_h_glmnb}\\
\ln r_{ij} &=& \beta_1 \ln H'_i H_j + \beta_2 C_{ji} + \beta_3 g_{ij} + \beta_4 g_{ji} \nonumber \\
                  & & + \beta_5 L_i + \beta_6 L_j + \beta_7 m_i + \beta_8. \nonumber
\end{eqnarray}
This model is intended to explain the number of help requests from $i$ to $j$ ($H_{ij}$) by the number of help requests from $j$ to $i$ ($H_{ji}$), the frequency of messaging $g_{ij}$ and $g_{ji}$, both players' levels $L_i$ and $L_j$, and their group size $m_i$. In addition, we used the log of the product of $H'_i$ and $H_j$, which respectively shows the total number of help requests from others to $i$ and from $j$ to others, because this value is expected to increase $H_{ij}$ proportionally, if players respond to help requests at random. 
That is, this model estimates the effects of these explanatory variables on $H_{ij}$ comparing the random behavior.
${\rm NB}(x)$ shows that $x$ follows a negative binomial distribution.
We estimated its parameters using $30,000$ relationships between players who satisfied $H'_i, H_j > 0$, sampled at random. 
We considerd this model and a GLMM with a Poisson distribution, because the data showed over-dispersion when the GLM with a Poisson distribution was applied, then AIC selected this model.

\begin{table}
\caption{Results of the regression analysis for response to help requests.}
\label{tbl_l2r_h_glmnb}
\begin{center}
\begin{tabular}{l|rl}
  Explanatory Variable & Regression Coefficient &（Standard Error） \\ \hline \hline
  $\ln H'_i H_j$  & $ 0.8891775$ & $ (0.0064618) ^{***}$ \\ \hline
  $C_{ji}$        & $ 0.0066710$ & $ (0.0018601) ^{***}$ \\ \hline
  $g_{ij}$        & $ 0.0089753$ & $ (0.0007516) ^{***}$ \\ \hline
  $g_{ji}$        & $ 0.0066472$ & $ (0.0007328) ^{***}$ \\ \hline
  $L_i$           & $ 0.0063142$ & $ (0.0002365) ^{***}$ \\ \hline
  $L_j$           & $-0.0046951$ & $ (0.0002432) ^{***}$ \\ \hline
  $m_i$           & $-0.0692416$ & $ (0.0012895) ^{***}$ \\ \hline
  Intercept       & $-0.4414666$ & $ (0.0218464) ^{***}$ \\ \hline
\end{tabular}
\end{center}
\end{table}

Table \ref{tbl_l2r_h_glmnb} shows the analysis results of the model.
The regression coefficient of $C_{ji}$, $g_{ij}$ and $g_{ji}$ were positive, even after controlling for $H_j$, $H'_i$, $L_i$ and $L_j$, i.e., in order to get helps from others, it was important to cooperate with each other and to communicate each other.
Additionally, this trend was stronger in smaller groups because $m_i$ was negative.
Thus, players selected partners who were cooperative for them even in the win--win situation.

\section{Discussion}

In the present study, we analyzed social behavior in the SNG to understand reciprocal altruism, one of the mechanisms that generate cooperative behavior. We showed the presence of reciprocal relationships with players who had reciprocal relationships receiving more benefits than others. That is, this SNG could be used to explore the mechanism of the emergence of reciprocal altruism.

Our analysis showed that cooperation and communication were important in creating reciprocal relationships. Cooperation was a signal representing cooperativeness (i.e., ``I'm a cooperator'') \cite{Andre2010}. 
A player needed to cooperate to send a signal. Consequently, players paid the cost of signaling, thereby providing benefits to receivers. The signal is expected to be highly reliable, because the meaning of the signal depends strongly on the sending method (cooperation) \cite{Smith1994a}.
Conversely, communication in simple text messages\footnote{Players tended to convey enthusiasm, acknowledgement, and encouragement.} was not guaranteed to be reliable as a signal, because signaling cost was not incurred, and it provided little benefit to receivers, i.e., with such a low-cost signal, senders could lie at any time\cite{Smith1994a}. Interestingly, low-cost signals not guaranteed to be reliable emerged, along with reliable signals.

Human social grooming is one of the low-cost signals. Social grooming is the construction and maintenance of social relationships in a complex society\cite{Kobayashi2011}. Apes groom each other as their social grooming\cite{Nakamura2003}. However, this is too costly for humans, because human group size is larger than ape group size; therefore, humans must invest time and effort in grooming others to create social relationships in large groups\cite{Dunbar2000}. Therefore low-cost social grooming (e.g., gaze grooming\cite{Kobayashi1997}) and one-to-many grooming (e.g., gossip\cite{Dunbar}) would be expected to have evolved in humans. In the SNG too, low-cost signals (``messaging'') might be used to create social relationships between players, even though a reliable signal was used.

Similarly, as the above reciprocal behavior, players more frequently tended to respond to requests from cooperative players with them than to those from noncooperative players, even in a win-–win situation, in which responding to a ``help'' request from others provides benefits to both the helper and the helped players. That is, cooperators acquired benefits for interacting with each other in such a situation. Conversely, noncooperators did not, because their ``help'' requests were received a lower priority than the cooperators' requests. Consequently, the cooperators acquired greater benefits than the noncooperators. We can regard the effect of such cooperators' behavior tendencies as punishment\cite{Rand2011} for noncooperators that may have decreased noncooperative behavior.

The present study provides quantitative evidence that the three mechanisms, reciprocal altruism, social grooming, and punishment, generate cooperation in an environment in which players can behave with fewer restrictions than in the theoretical and experimental studies.
In addition, this trend was stronger in smaller groups.

We regarded the social environment of the SNG as static for simplification. However, in reality, new players come to this game every day, some players stop playing, and new social relationships are constructed. Simultaneously, the strength and kinds of social dilemmas change dynamically. Therefore, analysis of the dynamics of social relationships is a challenge for future research.

\bibliographystyle{plain}

\begin{thebibliography}{10}

\bibitem{Andre2010}
Jean~Baptiste Andr\'{e}.
\newblock {The Evolution of Reciprocity: Social Types or Social Incentives?}
\newblock {\em The American Naturalist}, 175(2):197--210, February 2010.

\bibitem{Axelrod}
Robert Axelrod.
\newblock {\em {The Evolution of Cooperation: Revised Edition}}.
\newblock Basic Books, 2006.

\bibitem{Bainbridge2007}
William~Sims Bainbridge.
\newblock {The Scientific Research Potential of Virtual Vorlds}.
\newblock {\em Science}, 317(5837):472--476, July 2007.

\bibitem{Barkow1995}
Jerome Barkow.
\newblock {\em {The Adapted Mind: Evolutionary Psychology and the Generation of
  Culture}}.
\newblock Oxford University Press, October 1995.

\bibitem{Bshary2006}
Redouan Bshary and Alexandra Grutter.
\newblock {Image Scoring and Cooperation in a Cleaner Fish Mutualism}.
\newblock {\em Nature}, 441(7096):975--978, June 2006.

\bibitem{Castronova2006}
Edward Castronova.
\newblock {On the Research Value of Large Games: Natural Experiments in Norrath
  and Camelot}.
\newblock {\em Games and Culture}, 1(2):163--186, April 2006.

\bibitem{Dunbar2000}
Robin Dunbar.
\newblock {On the origin of the human mind}.
\newblock In P~Carruthers and A~Chamberlain, editors, {\em The Evolution of
  Mind}, pages 238--253. Cambridge University Press, 2000.

\bibitem{Dunbar}
Robin Dunbar.
\newblock {Gossip in Evolutionary Perspective}.
\newblock {\em Review of General Psychology}, 8(2):100--110, 2004.

\bibitem{Fehr2003}
Ernst Fehr and Urs Fischbacher.
\newblock {The Nature of Human Altruism}.
\newblock {\em Nature}, 425(23):785--791, 2003.

\bibitem{Grujic2010}
Jelena Gruji\'{c}, Constanza Fosco, Lourdes Araujo, Jos\'{e} Cuesta, and Angel
  S\'{a}nchez.
\newblock {Social Experiments in the Mesoscale: Humans Playing a Spatial
  Prisoner's Dilemma}.
\newblock {\em PLoS ONE}, 5(11):e13749, November 2010.

\bibitem{Grujic2012}
Jelena Gruji\'{c}, Torsten R\"{o}hl, Dirk Semmann, Manfred Milinski, and Arne
  Traulsen.
\newblock {Consistent Strategy Updating in Spatial and Non-Spatial Behavioral
  Experiments Does Not Promote Cooperation in Social Networks}.
\newblock {\em PLoS ONE}, 7(11):e47718, November 2012.

\bibitem{hauert2002}
Christoph Hauert, Silvia {De Monte}, Josef Hofbauer, and Karl Sigmund.
\newblock {Volunteering as Red Queen Mechanism for Cooperation in Public Goods
  Games}.
\newblock {\em Science}, 296(5570):1129--1132, May 2002.

\bibitem{Kobayashi2011}
Hiromi Kobayashi and Kazuhide Hashiya.
\newblock {The gaze that grooms: contribution of social factors to the
  evolution of primate eye morphology}.
\newblock {\em Evolution and Human Behavior}, 32(3):157--165, May 2011.

\bibitem{Kobayashi1997}
Hiromi Kobayashi and Shiro Kohshima.
\newblock {Unique Morphology of the Human Eye}.
\newblock {\em Nature}, 387(6635):767--768, June 1997.

\bibitem{Lindgren1991}
Kristian Lindgren.
\newblock {Evolutionary Phenomena in Simple Dynamics}.
\newblock In {\em Artificial Life II}, pages 295--312, 1991.

\bibitem{Nakamura2003}
Michio Nakamura.
\newblock {`Gatherings' of social grooming among wild chimpanzees: implications
  for evolution of sociality}.
\newblock {\em Journal of Human Evolution}, 44(1):59--71, January 2003.

\bibitem{Nowak2006}
Martin~A Nowak.
\newblock {Five Rules for the Evolution of Cooperation}.
\newblock {\em Science}, 314(5805):1560--1563, 2006.

\bibitem{Nowak1993}
Martin~A Nowak and K~Sigmund.
\newblock {A Strategy of Win-Stay, Lose-Shift That Outperforms Tit-for-Tat in
  the Prisoner's Dilemma Game}.
\newblock {\em Nature}, 364(6432):56--58, July 1993.

\bibitem{PACKER1977}
Craig Packer.
\newblock {Reciprocal Altruism in Papio Anubis}.
\newblock {\em Nature}, 265(5593):441--443, February 1977.

\bibitem{Rand2011}
David~G Rand, Samuel Arbesman, and Nicholas Christakis.
\newblock {Dynamic Social Networks Promote Cooperation in Experiments with
  Humans}.
\newblock {\em Proceedings of the National Academy of Sciences},
  108(48):19193--19198, November 2011.

\bibitem{Rand2013}
David~G Rand and Martin~A Nowak.
\newblock {Human Cooperation}.
\newblock {\em Trends in Cognitive Sciences}, 17(8):413--425, August 2013.

\bibitem{Rand2009}
David~G Rand, Hisashi Ohtsuki, and Martin~A Nowak.
\newblock {Direct Reciprocity with Costly Punishment: Generous Tit-for-Tat
  Prevails}.
\newblock {\em Journal of theoretical biology}, 256(1):45--57, January 2009.

\bibitem{Smith1994a}
John~Maynard Smith.
\newblock {Must reliable signals always be costly?}
\newblock {\em Animal Behaviour}, 47(5):1115--1120, May 1994.

\bibitem{Smitha}
John~Maynard Smith and E\"{o}rs Szathm\'{a}ry.
\newblock {\em {The Origins of Life: From the Birth of Life to the Origin of
  Language}}.
\newblock Oxford University Press, 2000.

\bibitem{Szell2012}
Michael Szell, Roberta Sinatra, Giovanni Petri, Stefan Thurner, and Vito
  Latora.
\newblock {Understanding Mobility in a Social Petri Dish}.
\newblock {\em Scientific Reports}, 2:457, January 2012.

\bibitem{Szell2010}
Michael Szell and Stefan Thurner.
\newblock {Measuring Social Dynamics in a Massive Multiplayer Online Game}.
\newblock {\em Social Networks}, 32(4):313--329, October 2010.

\bibitem{Szell2013}
Michael Szell and Stefan Thurner.
\newblock {How Women Organize Social Networks Different from Men}.
\newblock {\em Scientific Reports}, 3:1214, January 2013.

\bibitem{Tanimoto2007}
Jun Tanimoto and Hiroki Sagara.
\newblock {Relationship between dilemma occurrence and the existence of a
  weakly dominant strategy in a two-player symmetric game}.
\newblock {\em Biosystems}, 90(1):105--114, 2007.

\bibitem{trivers1971}
Robert~L. Trivers.
\newblock {The Evolution of Reciprocal Altruism}.
\newblock {\em Quarterly Review of Biology}, 46:35--37, 1971.

\bibitem{Wilkinson1990}
Gerald Wilkinson.
\newblock {Food Sharing in Vampire Bats}.
\newblock {\em Scientific American}, 262(2):76--82, February 1990.

\end{thebibliography}

\appendix

\section{Game Information}

We analyzed cooperative behavior in the SNG, ``Girl Friend BETA.'' Table \ref{table_target_service} presents the game information.

In this SNG, players create individual decks of cards that they collect and then use their decks to perform tasks in the SNG. A powerful deck, constructed from powerful cards, provides an advantage for game play in various situations. The players' primary motivation in the SNG is to obtain powerful cards. Players can obtain powerful cards as top-ranking rewards (see details later) or by casting lots called ``Gacha.''

Players belong to groups in which they cooperate with each other to play the game efficiently; the groups were limited to 1–-50 players. We filtered out players who do not belong to groups, because almost all active players belonged to groups to play effectively. Active players can create groups on their own. Others apply to join groups and then join a group after acceptance of the application by an administrator, who is typically a group founder. Players can leave a group at any time and apply to join a different group. Players observe their group members' behavior (e.g., attack on common enemies (see details later)), because the game system shows their behavior on the game screen. Thus, the SNG meets conditions 1 and 2 for reciprocity.

This game system provides two communication methods, ``messaging'' between two players and posting on a ``bulletin board,'' which is provided for each group. We analyzed ``messaging,'' because it was used more often than the ``bulletin board.'' Players use this function to send simple messages limited to 30 Japanese characters. It does not affect either senders or receivers negatively; nevertheless, its positive effects are also few. Every player can read messages of all other players on the respective receiver's profile page at any time. It is used primarily one to one, but players can also send a message concurrently to their group members or to players who have joined a battle with them (see details later).

\begin{table}
 \caption{Game information}
 \label{table_target_service}
 \begin{center}
  \begin{tabular}{c|c}
   Developer and Publisher & CyberAgent Inc. \\
   \hline
   Service Name & Girl Friend BETA \\
   \hline
   URL & http://vcard.ameba.jp \\
   \hline
   Event Name & Cherry Blossom Viewing Party \\
   \hline
   Event Type & Raid Battle\\
   \hline
   Event Time Period & 3/25/2013 16:00 to 4/8/2013 14:00 \\
   \hline
   Analysis Time Period & 4/5/2013 0:00 to 4/8/2013 14:00 \\
  \end{tabular}
 \end{center}
\end{table}

\section{Game Rules}
\begin{figure}
 \begin{center}
  \includegraphics[width=80mm,clip]{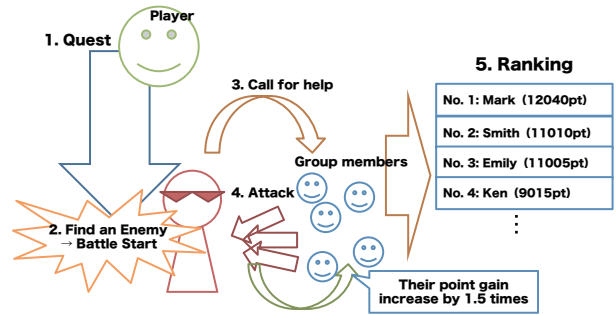}
 \end{center}
 \caption{Overview of raid event. A player conducts ``quests'' to find enemies (1). The player begins a battle upon finding an enemy and then attacks the enemy to obtain points (2). Enemies with very high hit points are strong; thus, they can call for help from other group members whom they have helped to win the battle (3). Players who helped had their point gain increased by $1.5$ times (4). Players compete in rankings based on their points (5).}
 \label{fig_raid}
\end{figure}

Our analysis target was a raid event (Fig. \ref{fig_raid}), in which players attack large enemies\footnote{
The enemy only has hit points as an attribute, meaning that players cannot be attacked by enemies. A player must attack an enemy to acquire event points at the expense of attack points.} and acquire ``event points.'' Players competed in the rankings based on their event points, because they received better awards as their rankings increased.

Players conduct quests\footnote{
This is one of the basic actions in SNGs. 
A player may encounter an enemy on performing certain action.} 
to find enemies during an event. Players begin battles when they find an enemy and then attack the enemy to obtain points. However, enemies with very high hit points are strong, making it difficult for players to win these battles unaided. Thus, they can call for help from other group members, to win the battle. Players who helped had their point gain increased by $1.5$ times. Therefore, players help their fellow group members to acquire more points.

Players' point gains are proportional to the amount of damage caused during attacks, i.e., more powerful decks earn more event points. A player immediately acquires points upon attacking an enemy, even if the enemy is not defeated. However, a player cannot battle another enemy while already battling another enemy, and that enemies' hit points increase with each battle; therefore, players must attack enemies repeatedly in the latter half of an event. Thus, a player who finds an enemy or helps a fellow group member must defeat the enemy before taking a next action, or wait until that the enemy leaves\footnote{
The length of the disable time is set between one and two hours. It is too long to complete the rankings for middle- and higher-rank players, because other players progress in the rankings during their disabled time.
}.

Players increase the amount of damage caused during their attacks by launching ``combo attacks,'' alternate attacks by two or more players in which the players need to launch attacks within ten minutes after other players\footnote{
If a player sequentially attacks an enemy then the attack is not count for the ``combo attacks.'' 
In addition, if players do not attack during ten minutes then their chain of combo attacks are reset to $0$.}.
The longer a chain of combo attacks, the more acquisition points are acquired. Battling enemies together with fellow group members increases the effectiveness of acquisition points.

Players must use a quarter of their attack points to attack; thus, they can attack four times when their point totals are full. There are two methods for replenishing these points: wait for the points to replenish over time or use an item that costs 100 JPY (such items are also sometimes distributed in the game as rewards).

Thus, players must use their resources (items and time) effectively to progress to a higher ranking, e.g., responding to a ``help'' request from their group members to acquire a point gain increase of $1.5$ times, increasing the number of ``combo attacks'' to increase the amount of damage, and reducing the disable time. We defined payment efficiency as the event points per payment, as in game theory.

\section{The Test Scenario}

It was impossible to track every cooperative behavior, because players can exhibit various behaviors in the SNG. Hence, we focused on one easily tracked cooperative behavior, and we regarded its frequency as players' cooperativeness.

We focused on the following scenario based on these rules to define players' cooperativeness.

\begin{itemize}
\item An enemy is attacked by a player and fellow group members.
\item The enemy's hit points are very few.
\end{itemize}

In this scenario, players who defeat the enemy will acquire only a few event points, because their attack power is higher than the enemy's hit points. Thus, their behavior is not efficient for acquiring event points. By contrast, if the players' attack power is lower than the enemy's hit points, their behavior is efficient for acquiring event points. Furthermore, they cannot battle another enemy, if battle with one enemy is ongoing, and therefore must wait until they defeat the enemy to exhibit efficient behavior.

\begin{table}
 \caption{
 Payoff matrix for the test scenario consisting of two players and an enemy with very few hit points. The player who attacks the enemy receives $S$, and the other player receives $T$. If neither player attacks the enemy, then each receives $P$. Attack by both players is impossible, because either player can defeat the enemy.
  }
 \label{table_payoff_matrix}
 \begin{center}
  \begin{tabular}{c||c|c}
    & Attack & Wait \\
   \hline \hline
   Attack & -, - & $S, T$ \\
   \hline
   Wait & $T, S$ & $P, P$ \\
  \end{tabular}
 \end{center}
\end{table}

In simple terms, consider that two players battled an enemy in this scenario, where their relationship is represented in Table \ref{table_payoff_matrix}. 
The relationship between the variables is $T> S > P$ in this payoff matrix. Attack is not efficient, when $S$ is less than $T$. 
However, if they do not attack the enemy, they waste time by waiting for someone else to attack, i.e., $P$ is lowest. 
It is not possible to cooperate both players in this scenario, because an attack on the enemy by either player immediately defeats the enemy. 
The values of this payoff matrix depend on each players situation, e.g., the differences between the two attack powers\footnote{In addition, it does not mean that the relationship between the payoffs is constant. If a player is about to go to sleep, then $S$ is larger than $T$, because the attack points replenish the next morning.}. 
This asymmetric diversity satisfies condition 3 for reciprocity. 
In the scenario, both try to avoid the worst situation (i.e., they get $P$), but they also do not want to pay the cost to avoid the worst situation (i.e., they get $S$). This social dilemma is similar to the one in the ``Leader game'' (Table \ref{table_leader_payoff_matrix}). In that game, Pareto efficiency is achieved when one cooperates, and the other does not. Then, the cooperator receives $S$, and the noncooperator $T$. 
That is, players receive a high payoff by sharing $S$ and $T$ on repeated plays of the game, a process known as $ST$ reciprocity\cite{Tanimoto2007}. 
We recognized this cooperative behavior, which provided the payoff $S$ from one to the other, as a cooperative behavior in this scenario.

Cooperative behavior is an inefficient attack, as shown in Table \ref{table_payoff_matrix}; thus we define $a_{ij}$ as the attack efficiency indicator:
\begin{eqnarray}
a_{ij} = e_{ij} / M(\bm{e_{i}}),
\end{eqnarray}
where $e_{ij}$ are the event points in player $i$'s $j$th attack and $M(\bm{e_{i}})$ is the median of $\bm{e_i} = \{e_{i1}, \cdots, e_{iN}\}$ ($N$ is the frequency of player $i$'s attacks). We considered cooperative behavior to be in the range of $a \leq 0.40$.

\section{The Test Scenario without Social Dilemma}

We analyzed social behavior in a win–-win situation in which players do not have a social dilemma, i.e., an enemy's hit point remains sufficient. It is difficult for players to defeat the enemy alone in this situation. Therefore, they request ``help'' from their group members. If the group members respond to the request, then the helper and the helped players effectively acquire event points by creating a chain of combo attacks. In addition, the group members also have their point gain increased by 1.5 times. Thus, responding to a ``help'' request from group members provides benefits to helpers, irrespective of the one they help.

\end{document}